%% file: Physrevised.tex
\documentclass[aps,prd,preprint,floatfix]{revtex4} 
\usepackage[dvips]{graphicx}
\usepackage{amssymb}
\newcommand{\be}{\begin{equation}}
\newcommand{\ee}{\end{equation}}
\newcommand{\bea}{\begin{eqnarray}}
\newcommand{\eea}{\end{eqnarray}}
\newcommand{\bitem}{\begin{itemize}}
\newcommand{\eitem}{\end{itemize}}

\newcommand{\benum}{\begin{enumerate}} 
\newcommand{\eenum}{\end{enumerate}}
\newcommand{\bc}{\begin{center}}
\newcommand{\ec}{\end{center}}
\newcommand{\omq}{\Omega_Q}
\newcommand{\ombe}{\Omega_B}

\newcommand{\dub}{w_Q}

\newcommand{\rhoq}{\rho_Q}
\newcommand{\rhob}{\rho_B}

\begin{document}
\title{Tracking Quintessence Would Require Two Cosmic Coincidences}

\author{Sidney Bludman}\email[]{bludman@mail.desy.de}
\affiliation{Deutsches Elektronen-Synchrotron
DESY,Hamburg}
\altaffiliation{University of Pennsylvania, Philadelphia}
\date{\today}

\begin{abstract}
  
  Good tracking requires that the quintessence energy fraction slowly
  increase while the roll $\lambda\equiv -d\ln V/\varkappa d\phi$
  slowly decreases, but is not yet truly slow-rolling.  The supernova
  bound on the present quintessence equation of state requires either
  (1) a cosmological constant or other fine-tuned "crawling
  quintessence" or (2) "roll-over quintessence" that tracked until
  recently, but now became slow rolling, because of a sharp increase
  in potential curvature.  Thus, fine-tuning is required by constant
  equation of state and inverse power potentials, but can be avoided
  by the SUGRA and Skordis-Albrecht potentials and other good
  trackers, provided quintessence energy domination and slow roll {\em
  both} began only recently.  This makes the time in which we live
  special in {\em two} respects.
\end{abstract}

\maketitle

\section{THE DARK ENERGY DENSITY IS NOW EXACTLY OR NEARLY STATIC}

The most surprising and important recent discovery in cosmology is
that the present universe is flat and dominated by unclustered dark
energy, whose energy density $\rhoq$ is now small and constant or
nearly constant.  Indeed, the next few years should show whether this

energy density is now truly static (cosmological constant) or slowly
evolving with the logarithm of the cosmological scale factor $N\equiv
\ln a=-\ln(1+z)$.  If dynamic, we would like to know something about the
dynamical equation of state $\gamma_Q(N)\equiv -d\ln
\rhoq/3 dN \neq 0$.  In this section, we review the present bounds on dark
energy evolution, before reviewing the constraints on {\em tracking
quintessence}, a particularly interesting model for the smooth
energy in terms of a very low mass canonical scalar field.

\subsection{Kinematics of the Accelerating Flat FRW Universe}

Supernovae Ia and cosmic shear directly explore the space-time
geometry by measuring the luminosity distance $d_L(z)=(1+z)\eta$
to individual distant supernovae, chosen to be standard candles,
and the angular-diameter distance $d_A(z)=\eta/(1+z)$ to distant
galaxies, from which the comoving distance \be\eta\equiv c\int_0^z
dz'/H(z')=c\int_0^t dt'/a(t')\ee  is inferred. This conformal
coordinate distance to the horizon, $\eta$, describes the proper
time evolution of the scale factor $a$.

The need for unclustered dark energy derives from general relativity,
which is tested to high precision in the classic solar system and
binary pulsar observations.  General relativity is also confirmed out
to cosmological distances and down to ages about one minute by BBN
light-element abundances, and is consistent with the observed spectra
of primordial density fluctuations, CBR temperature anisotropies, and
large-scale structure power.  We will therefore ignore alternate
gravity theories, such as small-curvature modifications of GR, large
extra spatial dimensions or brane
cosmology, and assume an homogeneous and isotropic
(Friedmann-Robertson-Walker) flat universe, the Friedmann expansion
rate is \be 8\pi G\rho=3 H^2 , \qquad \mbox{where} \quad H \equiv \dot
a/a. \ee Quantum field theory requires that the energy density, and
therefore $G$, be positive, so that we can write $\sqrt{8\pi
  G}\equiv\varkappa\equiv 1/M_P$, where $M_P=2.44e18~GeV$ is the
reduced Planck mass. The derived quantity, the cosmological fluid
pressure $P=-d (\rho c^2 a^3)/d a^3 $, may be positive or negative.

In terms of geometrical quantities, \be\varkappa^2 P/c^2=-(2\dot
H+3 H^2) ,\ee the enthalpy is \be \varkappa^2(\rho+P/c^2)=-2\dot
H=-d H^2/d N=-(\varkappa^2 /3)(d\rho/dN), \ee and the over-all
barotropic index is \be\gamma \equiv -d\ln\rho/3
dN=(\rho+P/c^2)/\rho= -\frac{2}{3}(d\ln H/dN).\ee Here the
logarithm of the cosmological scale factor $N\equiv \ln
a=-\ln(1+z)$, so that $dN=H dt$. This equation of state and its
quintessence component depends on the Hubble time $H^{-1}=d\eta/dz$,
and its derivative $d H^{-1}/d z=d^2\eta/dz^2$.

In flat FRW cosmology, the space-time curvature (Ricci scalar) is \be R\equiv -6(\dot H+2
H^2)=\varkappa^2 (3 P-\rho),\ee while the acceleration, $\ddot
a/a=-\varkappa^2 (\rho+3 P)/6$, so that \be \ddot a a/{\dot
a}^2=1- d(H^{-1})/d t=-(1+3 w)/2 \ee ranges from $-2$ to $1$, when
the overall equation of state $w\equiv P/\rho$ ranges from the
stiff value $1$ down to the ultra-soft value $-1$ . We now know
that, until about red-shift $z\sim 0.5$, attractive gravity
dominated the cosmological fluid so that large scale structures
could form. Only recently, after the expansion rate in equation
(7) outpaced the growth of the Hubble radius $H^{-1}$, did
$P/\rho=<-1/3$, and the expansion become accelerated.

\subsection{Present Cosmological Parameters}

Supernova observations were the first evidence for smooth energy and
remain
the direct observation of
acceleration of the present universe.  This supports the
possibility of an inflationary early universe, which is otherwise
not directly observable.

Because the barotropic index (5) and its quintessence component (13)
depend on the first and second derivatives of the comoving distance
$\eta$, the quintessence evolution $\dub(z)$ depends on first and
second derivatives of the observed luminosity distances \cite{Moar} or
on $H(z),~d H(z)/dz$. In
practice, quintessence is appreciable only for small red-shift. This
means that, before $\dub(z)$ can be extracted from the supernova data,
the inherently noisy luminosity distance $d_L(z)$ data must be
parametrized. For this, and other reasons, along with a large number
of high red-shift supernovae, precise knowledge of other cosmological
parameters will be needed \cite{Naka,Hut,Frieman}, and can still determine only one
or two parameters characterizing the potential, such as
$w_{Q0},~(d\dub/dz)_0$.  A worrisome feature $^{5)}$ of the present
supernova data is that, taken separately, the low $z<0.34$ and high
$z>0.34$ observations imply best-fit supernovae Ia absolute magnitudes
differing by 0.5 magnitude and are each compatible with an
Einstein-Lemaitre $\Omega_{Q0}=0$.  Indeed, the combined data actually
favors a supernova-averaged quintessence equation of state
\[\widetilde{\dub}(N)\equiv\int_N^0
 \dub(N')\,d N'/N <-1.\]

While programs to measure luminosity and angular diameter distances
are underway, we already know that we live at a time when
\be\Omega_{Q0}=0.71\pm 0.07 , \qquad \widetilde{\dub}<-0.78~(95\%~CL),
\qquad h\equiv H_0/100=0.72\pm 0.05 ,\ee that the radiation/matter
equality took place at red-shift $z_{eq}=3454^{+385}_{-392}$
\cite{Spergel,Tonry}, dark energy began dominating over matter
$\gtrsim 6.3~Gyr$ ago, and the cosmological expansion has been
accelerating since red-shift $z\sim 0.7$ \cite{Tonry,Blud}.  This fit
to $\widetilde{\dub}$ derives from SNIa, 2dFGRS and Ly$\alpha$
observations, and assumes $\dub$ constant and $>-1$.  Because the
observations average over a small range in $z$, in which $\dub (z)$
changes relatively little, $\widetilde{\dub}<-0.78$ is also a rough
upper bound to the present value $w_{Q0}$.

The radiation/matter background density, \be
\rho_B=(11.67 a+0.003378)/a^4\qquad meV^4 ,\ee is now
$\rho_{B0}=11.67 ~meV^4$ and was $\rho_{Bi}=0.003378~GeV^4$ at
fiducial red-shift $z=10^{12}$.  The supernova observations fit an
average
\[\widetilde{\dub}(N)\equiv\int_0^N \dub(N')\,d N'/N ,\]
The CBR anisotropy and mass fluctuation spectrum, on the other hand,
depend on "early quintessence" \cite{Cald}, back to the last
scattering surface $z\sim 1100$. Where needed, we will fix $h^2=1/2$,
so that the present critical density and smooth energy density are
$\rho_{cr0}=40.5~meV^4,~\rho_{Q0}=28.8~meV^4=(2.3~meV)^4$.

This completes our discussion of the geometry of an accelerated flat
universe.  The next two sections review scalar field dynamics and the
two Attractor Conditions, extending the many earlier optimistic
treatments of the Tracker Condition
\cite{CDS,SWZ,ZWS,Fer,Liddle,Cope,Ng} and of inverse power potentials
(Section IV) \cite{Ratra,Wett}, by considering (1) poor trackers, (2)
post-tracker behavior in the present quintessence-dominated era, (3)
the range of initial conditions leading to tracking, and (4) the
numerical problems encountered in cosmological dynamics, particularly
in the freezing and tracking epochs.  Sections IV and V conclude that the
nearly static upper bound (8) requires that any quintessence must be
either finely-tuned (crawling quintessence) or have large current
curvature (cross-over quintessence \cite{Wett2}) .

\section{DARK ENERGY AS A SCALAR FIELD}

\subsection{Quintessence Dynamics: Potentials Not Yet Truly Slow-Rolling}

Given that the GR universe is flat, presently dominated by smooth
energy, and recently accelerated, is this smooth energy constant (a
cosmological constant) or dynamic?  While there is certainly no
evidence for evolving smooth energy, attractive reasons for
considering it to be {\em dynamic} are: (1) Theoretically, it can
explain why the smooth energy density is now small but non-vanishing
(``Why small?), and can suggest reasons why the universe only recently
became smooth energy dominated and expanding (``Why now?''); (2)
Observationally, it may suggest the physical mechanism driving present
cosmological evolution.

The smooth energy may be due to tangled topological defects, Chaplygin
gases, trans-Planck or renormalization effects in ordinary curved
space-time, or due to scalar fields with generic \cite{Ratra,Wett}
attractor properties, that make the present smooth energy density
insensitive to a more-or-less broad range of unknown initial
conditions.  

The simplest dynamical realization for the smooth energy
is by  a spatially homogeneous light classical
scalar field, minimally coupled to gravity, with zero true
cosmological constant and kinetic energy.
{\em Canonical quintessence} assumes an effective Lagrangian linear in the
scalar field kinetic energy $K=\dot\phi^2/2$, rolling down its
self-potential $V(\phi)$ to zero.  (At the end of this section, we
briefly comment on k-essence, which assumes an effective Lagrangian
that is non-linear in the kinetic energy.)

In canonical quintessence, the scalar field equation of motion \be
\ddot\phi+3H\dot\phi+dV/d\phi=0,\ee has the first integral \be
\rhoq=\dot\phi^2 /2+V(\phi),\ee where \be V(N)=\rho_Q+d\rho_Q/6
dN=\rho_Q(1-\dub)/2,\qquad P/c^2=\dot\phi^2 /2-V(\phi).  \ee The
barotropic index \be \gamma_Q(N)\equiv -d\ln \rhoq/3
dN=((\rho+P/c^2)/\rho)_Q \equiv 1+\dub \ee lies between 0 and 2, so
that the Weak Energy Condition, $\rho+P/c^2>0, ~w>-1$, requires that
the density $\rho$ and the Hubble expansion decrease monotonically.
Because the scalar field does not cluster on supercluster scales, its
mass must be $\lesssim 10^{-31}~eV$.  Such an incredibly small mass is
hard to protect against matter-coupling and SUSY-breaking
interactions, unless the general relativity theory is expanded to a
scalar-tensor theory of gravitation.

From the energy integral (11), the
quintessence kinetic energy fraction of the total energy density
$x^2\equiv \dot\phi^2/2\rho=\gamma_Q \Omega_Q/2=
(\varkappa
d\phi/dN)^2/6$. The ratio of kinetic/potential energy
$K/V=(1+\dub)/(1-\dub)$ has the rate of change $
d\ln(K/V)/dN=6(\Delta-1)$, where $\Delta(N)\equiv -d\ln V/3\gamma_Q
dN$. Thus, the {\em roll}, \be\lambda\equiv -d\ln V/\varkappa d\phi=
\sqrt{3\gamma_Q/\Omega_Q}\cdot \Delta, \ee and \be d\dub/dN
=3(1-\dub^2)(\Delta-1), \qquad \varkappa d\phi/d
N=\sqrt{3\gamma_Q\Omega_Q}\ee is a two-element non-autonomous system
for the dependent variables $\phi,~\dub$. Integrating the second
equation (15) implicitly relates $\phi$ and $V(\phi)$, so that, if the
equation of state $\dub (z)$ can be observed, the potential can be
reconstructed. The overall equation of state is \be
w=\gamma-1=w_B\ombe+\dub\omq,\ee where the dimensionless ratios,
$\omq, ~\ombe\equiv \rhob/(\rho+\rhoq)$, are the energy density
fractions in quintessence and in the radiation/matter background, and
$\gamma_Q,~\gamma_B \equiv -d \ln \rhob /3 dN $ are their
corresponding barotropic indices.

Defining the potential energy fraction $y^2 \equiv V_Q/\rho$,
equations (15) have different {\em scaling solutions} whenever
$\dub$ or $K/V \equiv (x/y)^2 =\gamma_Q/(2-\gamma_Q) \approx const$:
{\em kination}, when $\dub\approx 1,x>>y$; {\em freezing}, when
$\dub \approx -1,x<<y$; {\em tracking}, when
$\Delta \approx 1,~y \propto x$.  The scale grows
as $a \sim t^{2/3\gamma_B},~ t^{2/3\gamma_Q}$, in the radiation/matter
background
and in the quintessence eras respectively.

Besides the roll $\lambda$, the potential is characterized by the
curvature \be \eta\equiv V''/V=\lambda^2-\lambda'=\lambda^2
\Gamma,\ee where \be\Gamma\equiv V V''/V'^2 \equiv \eta/\lambda^2 \equiv (1+1/\beta), \quad 1/\beta \equiv\Gamma-1=(1/\lambda)', \quad '\equiv
d/\varkappa d \phi . \ee When the roll is flat
($\epsilon\equiv\lambda^2/2 \ll 1$), the kinetic energy
$\dot\phi^2 /2$ is negligible in the quintessence energy (12).
When $\eta\ll 1$, $\ddot\phi$ is negligible in the equation of
motion (10).  In ordinary inflation, both these conditions hold
({\em slow roll approximation}): the expansion is dominated by the
cosmological drag, and the field is nearly frozen.

In quintessence, on the other hand, the acceleration began only
recently, when $\varkappa\phi_0\sim 1$, so that the roll
$\lambda_0$ and curvature $\eta_0$ are still $\mathcal{O}(1)$, and
their difference $\eta-\lambda^2=(\ln V)''=-\lambda'$ may be large or small,
and their ratio $1<\Gamma=\eta/\lambda^2$ may be large. This
invalidates the slow roll approximation for quintessence, so that
the dynamical equations need to be integrated numerically. We
ultimately handled the kination/freezing and freezing/tracking
transitions and numerical stability and round-off problems in the
protracted frozen era, by implicit Adams backward differentiation
procedures (Maple lsode[adamsfunc] and lsode[backfull]), with
small adaptive step-size.

Unless it undergoes a symmetry-breaking phase transition, the
quintessence potential rolls monotonically towards a minimum at
$\phi=\infty$ or at some finite $\phi_{min}$. A potential minimum
realizes a large potential curvature. Either way, the
potentials we consider are always convex. (Presumably, there is no
true cosmological constant so that the potential energy and the Hubble
expansion vanish asymptotically, avoiding the possibly worrisome
future de Sitter event horizon, with attendant future causal
isolation.)

Both the roll and the curvature, $\lambda$ and $\eta$ are listed in Table
I, for five different potentials. The third row gives the 
constant $\eta/\lambda^2$ or $\beta=const\equiv\alpha$ potential.
The first and fourth rows 
give its $\beta=0,~\infty$ limits, the cosmological
constant and exponential potentials.  On the second row,
where $\tilde{\alpha}\equiv\sqrt{\gamma_Q}/\alpha$, the unrealistic
{\em constant $\dub$ model} interpolates between the inverse power
potential, when $\tilde{\alpha}\varkappa\phi\ll 1$, and the
exponential when $\tilde{\alpha}\varkappa\phi\gg 1$.  The bottom row
is the more realistic inflaton broken-SUSY SUGRA potential,
in which, when  $\phi \gtrsim M_P $, $\beta(\phi)$ decreases and the
curvature $\eta(\phi)$ increases significantly

\subsection{Phaseportrait in Terms of Quintessence Kinetic, Potential
Energy Canonical Variables}

In place of the phase variables $\phi,~\dub\equiv (P/\rho)_Q ,$ we may
use $x,~y$ , for which the equations of motion
are \cite{Fer,Liddle,Cope,Ng} \bea
dx/dN &=& -3x + \lambda\sqrt{3/2}y^2 + 3x\gamma/2 \\
dy/dN &=&   \qquad    - \lambda\sqrt{3/2}xy+ 3y\gamma/2 \\
d\lambda/dN &=&-\sqrt 6 \lambda^2 x/\beta \qquad \mbox{or}\qquad
d(1/\lambda)/d N=\sqrt 6x/\beta.\eea \input{TableI.tex} The
overall equation of state of our two-component mixture of radiation/matter
background and quintessence, $\gamma=\gamma_Q\omq+\gamma_B\ombe=
2x^2+\gamma_B (1-x^2-y^2), $ is a time-dependent function of the
scalar field $\phi(N)$. Thus, \be x^2+y^2=\omq,\quad 2 x^2=\omq
\gamma_Q, \quad y^2/x^2=(1-\dub)/(1+\dub), \quad d\ln (x^2/y^2)/dN=6(\Delta
-1).\ee The three-element system (19-21)) is autonomous, except
for the slow change in $\gamma_B(N)$ from 4/3 to 1, while
gradually going from the radiation-dominated to the
matter-dominated universe, around red-shift $z_{eq}=3454$ .

The magnitude of $V$ needs to be fine-tuned to the present value
$V_0=\rho_{cr0}y_0^2=\rho_{Q0}(1-w_{Q0})/2$. For example, inverse
power potentials, require the energy scales $M_{\alpha}=(V_0\phi_0
^\alpha)^{1/(4+\alpha)}$, listed in the third column of Table 2.
For shallow potentials ($\alpha<0.2$), this energy scale is close
to observed neutrino masses and to the present radiation
temperature, possibly suggesting some role for the neutrino mass
mechanism or for the matter/radiation transition, in bringing
about quintessence dominance. For steep potentials ($\alpha>1$),
this mass scale can be considerably larger, suggesting the larger
scales we encounter in particle physics.

While the evolution of a homogeneous scalar field depends only on its
equation of state $\dub=(P/\rho c^2)_Q$, the growth of its
fluctuations depends also on the quintessence sound speed $c_s^2=(d
P/d \dot\phi)/(d\rhoq/d \dot\phi)$. With the linear form for the
kinetic energy $K=\dot\phi^2/2$ that canonical quintessence assumes,
$c_s^2=c^2$ and $-1\leq \dub\leq 1,~d \dub/dz>0$. Non-canonical scalar
fields, such
as k-essence \cite{Armen,Erikson}, permit $\dub<-1,~d
\dub(z)/dz<0$ and give different sound speed and structure evolution.

Effective actions with such non-linear kinetic energy appear in some
string and supergravity theories.
Despite this possible difference in sign of $d\dub/dz$, unless
$c_s^2\sim 0$ since the surface of last scattering, k-essence is
hardly distinguishable from quintessence \cite{Barger}. Although
tracking trajectories in the radiation-dominated era, are quickly
attracted towards $\dub\approx -1$ 
in the matter-dominated era, k-essence, like quintessence, 
requires fine-tuning. The most catastrophic consequence of
violating the Weak Energy Condition is that the scalar field energy
can grow with time, ultimately (after about 35 Gyr of scalar field
dominance \cite{CalKam}) ripping apart bound systems, such as
galaxies, planets and atoms.

\input{figures1.tex}

\input{TableIIrevised.tex}

\input{QuinTrack.tex}

\input{figures2.tex}
\input{InvPower.tex} \input{Conclusions.tex}

\begin{acknowledgments}
We are indebted to US
Department of Energy grant DE-FGO2-95ER40893 at the University of
Pennsylvania for partial travel support.

\input{bibliographyMykanos.tex}
\end{acknowledgments}
\end{document}

%% file: TableI.tex
\begingroup
\begin{table*}
\caption{Potentials described by roll $\lambda=-d\ln V/\varkappa
d\phi$ and curvature $\eta=d^2 V/V d(\varkappa \phi)^2$.}
\vspace{.2cm}
\scriptsize 
\begin{ruledtabular}

   \begin{tabular}{|l|c|c|c|r|}
   {\bf $V(\phi)$ }                       &{\bf $\lambda(\phi)$}
   &{\bf $\eta(\phi)=\lambda^2\Gamma$}&{\bf
     $\Gamma-1=1/\beta(\phi)$}&NAME
 \\  \hline
   $\exp{-\lambda\varkappa\phi}$
   &$\lambda=\sqrt{3\gamma_B/\Omega_Q}
   $ &$\lambda^2=const>3\gamma_B$        &$0$  &exponential \\
   $1/\sinh^{\alpha}(\tilde{\alpha}\varkappa\phi)$
   &$(\alpha\tilde{\alpha})\coth(\tilde{\alpha}\varkappa\phi)$
   &$(\alpha\tilde{\alpha})^2 [(1+\alpha)\coth(\tilde{\alpha}\varkappa\phi)-1]$
   &$1/\alpha\cosh^2(\tilde{\alpha}\varkappa\phi) $
   &$\dub=-2/(2+\alpha)$ \\
   $\phi^{-\alpha}$ &$\alpha/\varkappa\phi$
   &$\alpha(\alpha+1)/(\varkappa\phi)^2$     &$1/\alpha$
   &inverse power \\
   $const$ &0  &0 &$\infty$  &cosmological const \\
   $\phi^{-\alpha}\cdot\exp{\frac{1}{2}(\varkappa\phi)^2}$
   &$\alpha/\varkappa\phi-\varkappa\phi$
   &$[\alpha(\alpha+1)+(1-2 \alpha) (\varkappa\phi)^2+(\varkappa\phi)^4]/(\varkappa\phi)^2$
   &$(\alpha+(\varkappa\phi)^2)/(\alpha-(\varkappa\phi)^2)^2$ & SUGRA
   \end{tabular}
\end{ruledtabular}
\normalsize
\end{table*}
\endgroup

%% file: figures1.tex
\begin{figure}[t!] 
\scalebox{0.7}{\includegraphics{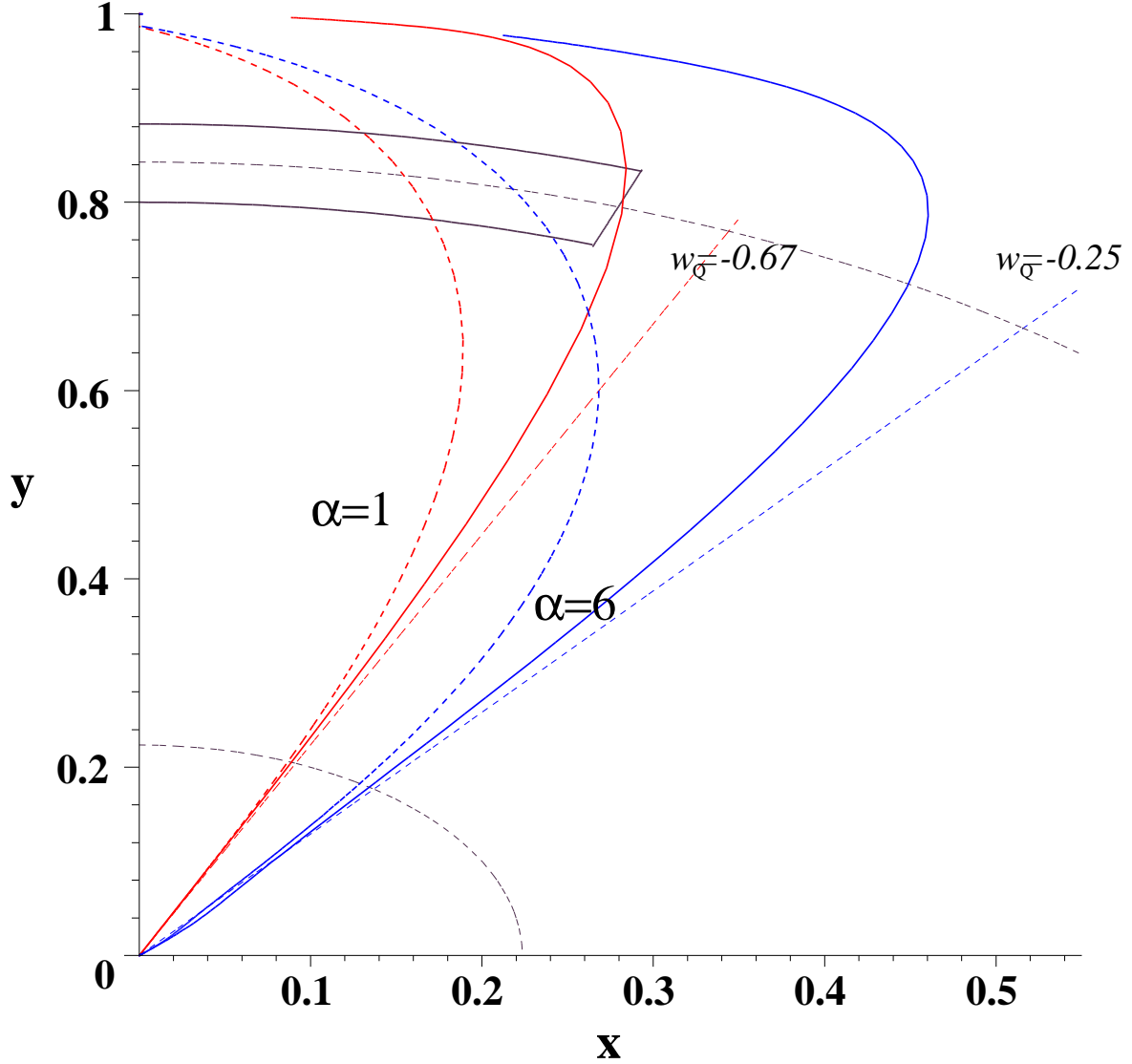}}
\caption{ Complete phaseportraits of instantaneous attractors, from
  red shift $z=10^{12}$ in lower left corner, to $z=10^{-10}$ on the
  top, now passing through $\Omega_{Q0}=0.71$. The evolution is shown,
  on the right for three steep potentials($\alpha=6$); on the left,
  for three shallow potentials ($\alpha=1$).  The evolution of each
  model is shown against the dashed circles $\Omega_Q=x^2+y^2=0.71$ at
  present, $=0.1$ when quintessence became significant.  In the
  background-dominated era $\Omega_Q \ll 1$ (lower left), all
  attractors track with $y\propto x$ and $w_{Qtr}=-0.25, -0.67$ for
  $\alpha=6,~1$ respectively. When quintessence begins dominating
  $\Omega_Q >0.1$, the 
  inverse power trajectories (solid), start late to curve slowly
  towards the y-axis ($w_Q=-1$); both SUGRA trajectories (dotted),
  start early to curve rapidly towards the y-axis.  The constant $w_Q$
  trajectories are shown dashed. The trapezoidal
  region on the upper left is the observationally allowed present
  phase space.}
\end{figure}

%% file: TableIIrevised.tex
\begin{table*}
 \caption{Tracker and present ($\Omega_{Q0}=0.71$) attractor solutions
   for inverse power and SUGRA potentials.}
 \begin{ruledtabular}
   \begin{tabular}{|r||c||c|c|c|c|c|c|c|c||c}
$\alpha$ &$w_{Qtr}$      &$M$
&$\varkappa\phi_0$&$\lambda_0$&$\eta_0$&$x_0^2$ &$y_0^2$&$w_{Q0}$
&$w_{Q1}$
  &$\Omega_{Qimax}$ \\  \hline

0     &  -1             &2.5~meV     & -   &0 (static)&0&0 &.71&-1
&0

                      &8e-45 \\
0.1   &  -0.937..-0.952 &12.1~meV    &0.285&0.351&1.36&.011
&.698&-0.97 &0.04

               &3e-36 \\
0.5   &  -0.733..-0.80  & 4.8~eV     &0.725&0.689&1.42&.049
&.661&-0.86 &0.08
                     &8e-10 \\ \hline
1     &  -0.555..-0.667 & 2.2~keV    &1.123&0.890&1.58 &.080
&.630&-0.78 &0.30
                   &3e-2  \\
6     &   0..-0.25      & 5.3~PeV    &4.029&1.489(fast)&2.59&.201
&.509 &-0.41 &0.40
              &3e-1 \\ \hline
1     &  -0.555..-0.667 & 2.1~keV    &0.870&0.280&2.40&.024
&.685&-0.82 &0.72
                      &3e-2  \\
6     &   0..-0.25      & 3.0~PeV    &2.254&0.408&2.35&.048
&.663&-0.87 &0.58
                     &3e-1 \\ 

   \end{tabular}
 \end{ruledtabular}
\end{table*}

%% file: QuinTrack.tex
\section{ATTRACTORS IN TRACKING AND IN QUINTESSENCE ERAS}

We will consider only canonical quintessence evolution with the
Friedmann expansion rate (2).The scalar field equations of motion
enjoy a fixed attractor solution $\lambda=const$ (exponential
potential), when $\beta=\infty$, and two different stable {\em instantaneous attractor}
solutions \cite{Wett,SWZ,ZWS,Cope}, when $\lambda$ is slowly varying in
equation (21), because $-\lambda'=\eta-\lambda^2\ll
1,~\beta\gg\lambda^2$.  Depending on the roll or $t\equiv
\lambda^2/3$, these are: The Slow-Roll (``Future'') Attractor 

\be x_c^2(\lambda)=t/2, \quad y_c^2(\lambda)=1-t/2, \quad
\Omega_{Qc}=1, \quad t<2 .\ee
The Fast-Roll (``Past'') Attractor
\be x_c^2(\lambda)=\gamma_Q^2/2t, \quad y_c^2(\lambda)=\gamma_Q
(2-\gamma_Q)/2t, \quad \Omega_{Qc}=\gamma_Q/t, \quad t>\gamma_Q , \ee
where
the equation of state \cite{Ng} is \small
\[
\gamma_{Qatt}(\lambda)=\frac{1}{2}[((1+2/\beta)t+\gamma_B)-\sqrt{((1+2/\beta)t-\gamma_B)^2+8\gamma_B
t/\beta}]=\left\{\begin{array}{ll} 0, &\beta=0 \\
                                        t, &t \ll 1 \\
                                        \gamma_B\beta/(\beta+2), &t \gg 1\\
                                        \gamma_B  ,&\beta=\infty .
                   \end{array} \right. 
\]
\normalsize 

Tracking quintessence \cite{Ratra,Wett,SWZ} exploits the Fast-Roll
Attractor to explain the very small present smooth energy density
dynamically (``Why small?''). Instead of a finely tuned cosmological
constant, quintessence requires a potential strength fine-tuned to the
present time (``Why now?'').  We reserve the term {\em tracker} for
these Fast-Roll Attractors in the radiation/matter
background-dominated era, for which $\beta$ and $\gamma_{Qatt}=2
\gamma_B/(2+\beta)$ are nearly constant. The Tracker Condition can
then be written as \be \Delta
=\lambda/\lambda_{tr}=\sqrt{\gamma_Q/\gamma_{Qtr}}\approx 1, \qquad
|d \ln \beta/dN|\ll 1 ,\ee where
$\lambda_{tr}\equiv\sqrt{3\gamma_{Qatt}/\omq}$.  The quintessence
fraction scales as 
$\Omega_{Qatt}\sim
a^{6 \gamma_B/(\beta+2)} \sim t^P$, where $P \equiv 4/(2+\beta)$. In
{\em good trackers} ($\beta\gg 1,~\gamma_{Qatt}\ll \gamma_B)$, $\omq$
and $\lambda_{att}$ vary slowly.  Indeed, very steep ($\beta \gg 1$)
potentials approximate exponential potentials and track for a very
long time, with slowly decreasing $\lambda$ and increasing $\omq$.
They show {\em early quintessence} because, from $\lesssim 0.05$ at
nucleosynthesis, $\omq (z)$ slowly increases and is already
non-negligible at recombination and structure formation.  In {\em poor
trackers} ($\beta<1,~\gamma_{Qtr}<2\gamma_B/3$), the roll is slow,
$\rho_{Qtr}$ is slowly decreasing (``crawling''), and $\omq\sim t^2,
~\lambda_{tr}\sim t^{-1}$ are changing with time.

At present, the roll is no longer fast and $\gamma_Q(\lambda)$ is
decreasing, curving the attractor trajectories towards the y-axis.  We
therefore need to study the quintessence-dominat post-tracking era
when $\Delta-1=d \ln (K/V)/6 dN<0, ~\dub\rightarrow -1$. Figure I
shows the instantaneous attractor phaseportraits, while tracking and
while quintessence-dominated, for two inverse power (solid) and two
SUGRA (dotted) potentials, all fitted to reach $\Omega_{Q0}=0.71$, at
the present time.  The constant $\dub=-0.67,-0.25$ trajectories are
shown dashed. For each form of potential, the shallow potential
$\alpha=1$ attractor trajectories appear on the left, the steep
potential $\alpha=6$ attractor trajectories on the right.  The
cosmological constant trajectory $\alpha=0$ is the y-axis.  The
exponential potential has fixed $\omq$; its fixed point,
$x=\sqrt{\Omega_{Q0}/2}=y$, would fall on the circle
$\Omega_{Q0}=0.71$, just off the right of the figure.

Along these trajectories, evolution is measured by the quintessence
fraction $\omq=x^2+y^2$, which is zero in the distant past (lower left
corner), unity in the far future (top of phaseportrait), and has been
fitted to $\Omega_{Q0}=0.71$ at the present time.  In the
background-dominated era, all phase trajectories track with nearly
constant slope $(y/x)_{tr}=\sqrt{V/K}$ gradually increasing from
$\sqrt{1/2+3/\beta}$ in the radiation-dominated era to
$\sqrt{1+4/\beta}$ in the matter-dominated era, and equation of state
$w_{Qtr}=-2/(\beta+2)$. Later, as quintessence grows, the three
potential forms curve differently towards the y-axis: the constant
$\dub$ potentials have been chosen to make their trajectories hold
their tracker values; the SUGRA potential
phase trajectories curve strongly towards the y-axis. The
presently-observed trapezoidal region in phase space nearly excludes
both inverse power potential, but comfortably allows the
SUGRA potential for a range in parameter $\alpha$.

The bottom half of Table II summarizes, on rows five and six, seven and
eight, the present properties of the $\alpha=6,1$ inverse-power and
SUGRA phase trajectories respectively.  (The top half of Table I also
includes, on row 2, the
cosmological constant phase trajectory, which is the y-axis in Figure 1.
Rows 3-4, summarize the nearly static $\alpha=0.5,0.1$ phase
trajectories, which are not plotted in Figure 1.)
For each $\alpha$, the second column tabulates the range in
tracker equation of state $w_{Qtr}$, running from the early
radiation-dominated era, to the late matter-dominated era.  To the
right of the double line, columns 3-6, give the potential energy scale
$M_\alpha$ needed to reach the present $\Omega_{Q0}=0.71$, and the
present values of the quintessence field, roll, and curvature.
Columns seven and eight give the quintessence kinetic and potential
energy fractions composing the total quintessence energy fraction
$\Omega_{Q0}$.  Column nine shows that only the high-curvature
SUGRA potentials give $w_{Q0}<0.78$, as observationally required.
If, in place of a cosmological constant,  quintessence
potentials like these exist, then the present red-shift variation
$w_{Q1}\equiv ((1+z) d \dub(z)/dz)_0$ in
column ten may ultimately be observable.
The last column gives the upper bound to the basin of attraction, to
be discussed in the next section.

%% file: figures2.tex
\begin{figure*}[t!]
\includegraphics*[bb=72 72 540 719,angle=-90,width=7.5cm]{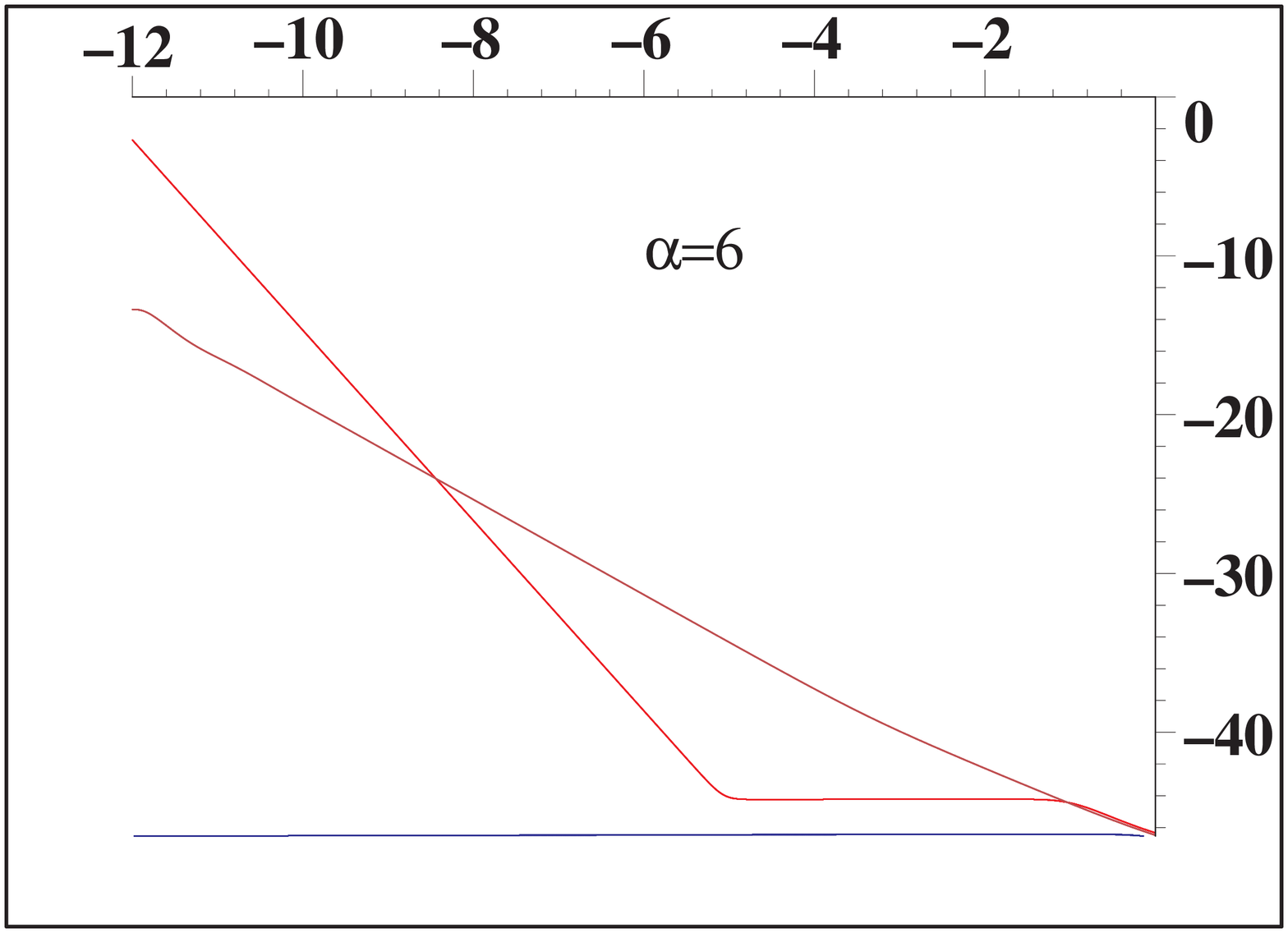}
\includegraphics*[bb=72 72 540 719,angle=-90,width=7.5cm]{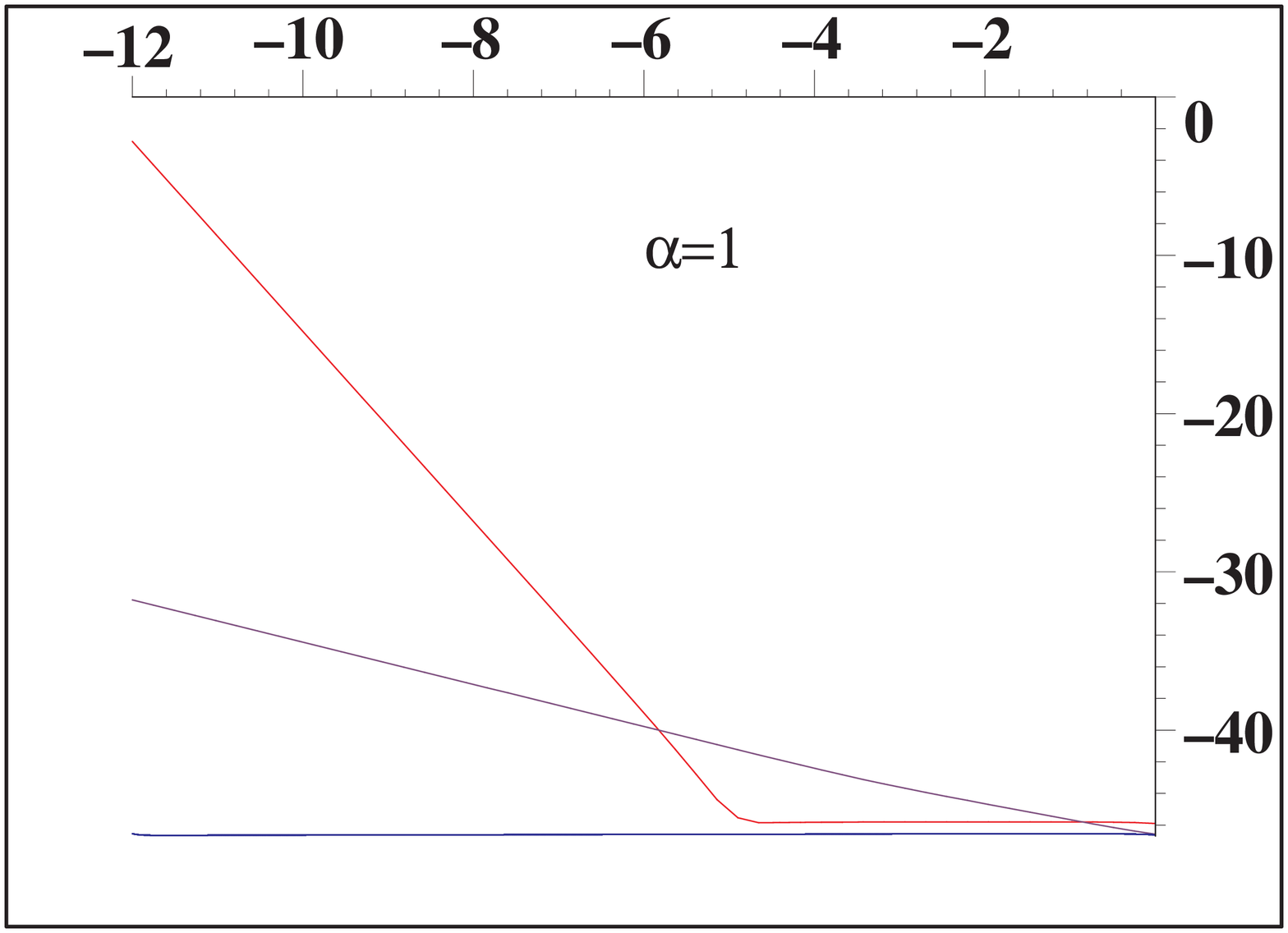}
\includegraphics*[bb=72 72 540 719,angle=-90,width=7.5cm]{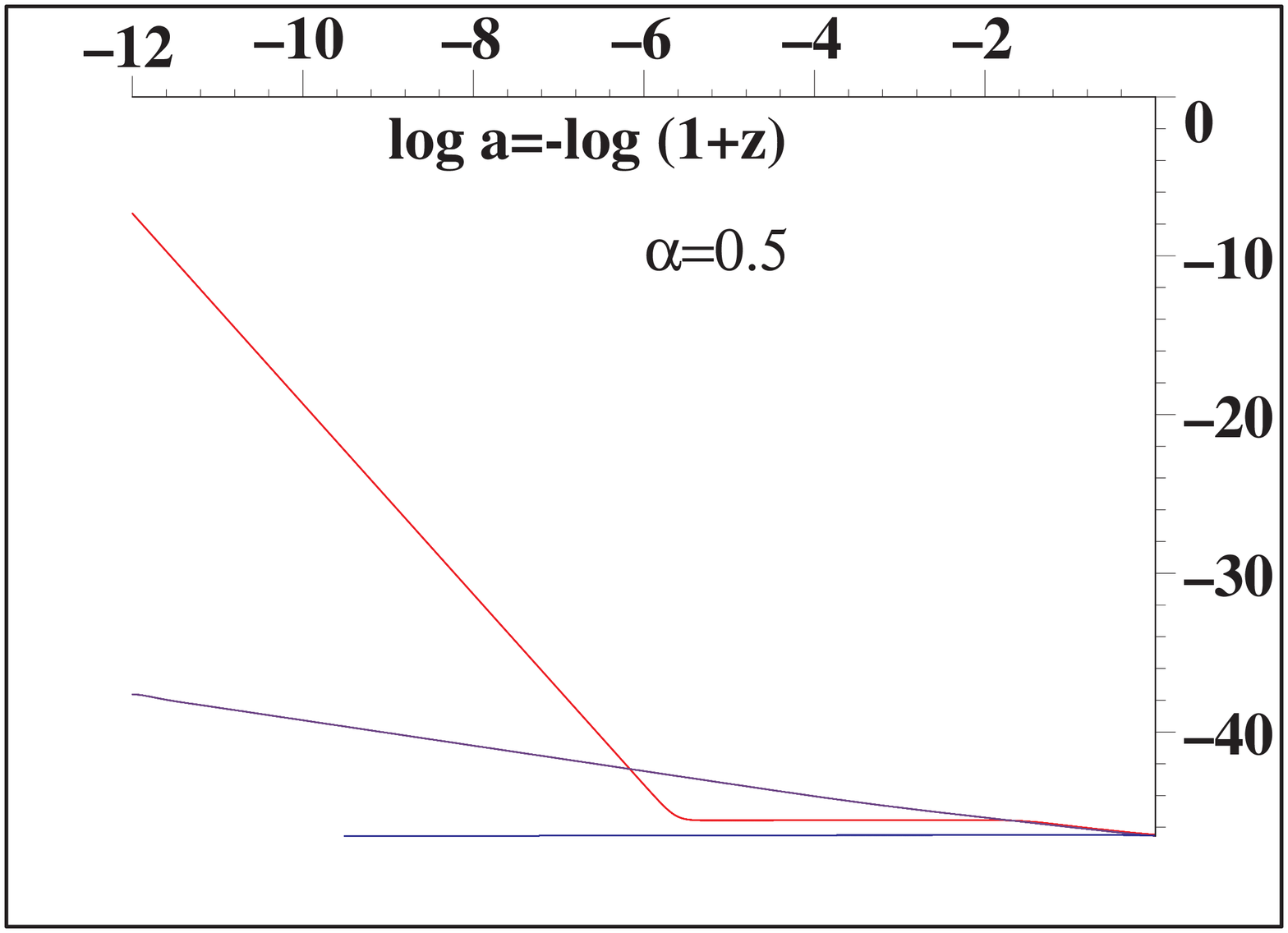}
\includegraphics*[bb=72 72 540 719,angle=-90,width=7.5cm]{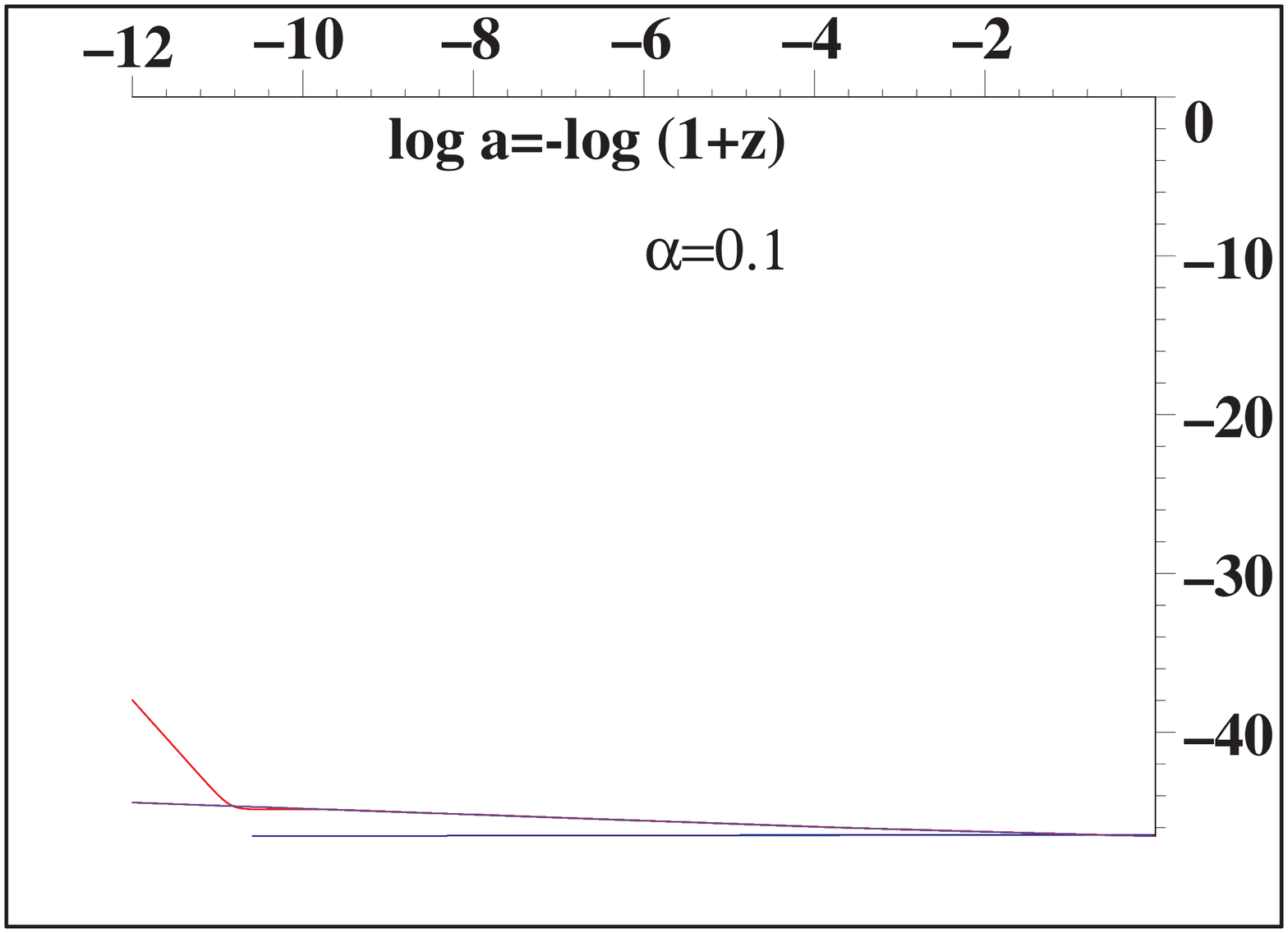}
 \caption{\label{Fig2.eps}
Evolution of quintessence energy density, log $\rho_Q (GeV^4)$, on
vertical axes, for four inverse power potentials
$\alpha=6,1,0.5,0.1$, from red-shift $z=10^{12}$, to the present
value $\rho_{Q0}=0.71\rho_{cr0}$. In all figures, the central,
trajectory is the attractor, starting with tracker slope $d
\ln\rho_Q/d N=-6\gamma_B/(2+\alpha)$. The lower curve is the
maximal undershoot trajectory, which freezes immediately and then
crawls slowly to join the attractor now. The upper curve is the
maximal overshoot trajectory, which kinates with slope -6, before
freezing late and now reaching the attractor.  Poor trackers 
freeze early, out of a narrow range in log $\rho_{Qi}$. }

\end{figure*}

%% file: InvPower.tex
\section{GOOD TRACKERS MUST NOW ROLL-OVER TO SLOW ROLL}
\subsection{Basin of Attraction for Trajectories Flowing onto the Tracker}

To obtain the presently small smooth energy without fine-tuning
initial conditions, phase trajectories must flow onto the
attractor before now and for a broad range of initial conditions,
the {\em basin of attraction}. A good tracker starts from a broad
basin of attraction and can still freeze relatively late,
before tracking with appreciable kinetic energy $K \lesssim V$.  A poor
tracker starts from a narrow
basin of attraction and freezes early and for a long time, before tracking with
small kinetic energy $K \ll V$.

Tracking is always preceeded by freezing, which begins when the
curvature drops, while the roll
is still appreciable.  From any {\em undershoot} initial conditions
$\rho_{Q0}<\rho_{Qi}<\rho_{Qtr}$, freezing starts directly. (Indeed,
if the initial density
 $\rho_{Qi}$ is small enough , the field crawls
down to its present value $\rho_{Q0}$, without ever tracking.) But,
starting from {\em overshoot} initial conditions
$\Omega_{Qi}>\Omega_{Qtr}$, freezing starts much later, at a value
$\phi_{fr}\gtrsim M_P \sqrt{6\Omega_{Qi}}$, only after a long kinated
era ($x \gg y$), during which $\dot{\phi}^2\approx 2\rhoq$ lets the
field grow rapidly while $\omq$ decreases. Above some
$\Omega_{Qimax}$, too much overshoot would make the phase trajectories
freeze so late as to reach the attractor only in the future.
We will now extend earlier treatments \cite{ZWS,SWZ} of trackers
to solutions that reach the
attractor only now, in the present, quintessence-dominated era.

\subsection{Inverse Power Potentials Do Not Show Enough Present Curvature}

For the simple inverse power law potentials \be
V(\phi)=M_{\alpha}^{4+\alpha}/\phi^\alpha,\ee 
$\beta(\phi)=constant\equiv\alpha$, so that the equation of motion (10) has
exactly scaling solutions in both the radiation- and the
matter-dominated eras. These much-studied potentials are interesting,
because they approximate any potential, while tracking. They arise naturally in supersymmetric
condensate models for QCD or instanton SUSY-breaking
\cite{Bin,Brax,Brax2}, but aquire appreciable quantum corrections
when $\phi\gtrsim M_P$. For these potentials, $\lambda\sim
V^{1/\alpha}\sim (yH)^{2/\alpha}$, the third equation (19)
integrates to $\lambda=\lambda_0(yH/y_0 H_0)^{2/\alpha}$ in terms
of present values of $y,~H,~\lambda$.

The second column in Table II gives the range in tracker values
$w_{Qtr}$, from $(\alpha-6)/3(\alpha+2)$, during the
radiation-dominated era, to $-2/(\alpha+2)$, during the
matter-dominated era.  The third column tabulates the quintessence
energy scales $M_\alpha$ needed, in order to fit
$\Omega_{Q0}=0.71$.

After tracking, these
trajectories curve towards the $x=0$ (asymptotic de Sitter)
axis. Between the two vertical double bars, columns four through nine,
summarize the present (post-tracker) values for trajectories that
have tracked before now. The steep $\alpha=6$ trajectory
would by now track down only to $w_{Q0}=-0.41$, which is excluded
observationally by the constraint (8), which requires
$\alpha\lesssim 1$ \cite{SWZ}. Inverse power trackers that are now approximately slow-rolling
($\lambda_0 < 1$) could only have been reached
from a narrow tracking basin of attraction. Indeed, in the static
limit (cosmological constant), the present smooth energy must arise
out of the unique initial condition $\rho_{Qi}=\rho_{Q0}$.

Integrating equations (17,18), the last column tabulates, for each
$\alpha$, the maximum initial quintessence fraction that would track
and finally reach the presently-allowed value $\Omega_{Q0}=0.71$.
Likewise, the top curves in each of Figures 2 show, on a logarithmic
scale, the maximum initial quintessence energy density that would
finally reach $\rho_{B0}=28.8 meV^4$. For
large $\alpha$, the quintessence potential would now still be
fast-rolling \cite{Blud}. As $\alpha$ decreases, $w_{Q0}$ decreases,
but the basin of attraction shrinks. For $\alpha<0.5$, the range in
initial values $\Omega_{Qi}$ that track before now is already almost
10 orders of magnitude narrower than the initial range of the good
$\alpha=6$ trackers first considered \cite{SWZ}. For a cosmological
constant ($\alpha=0$), the present vacuum energy is realized only if
it is initially tuned uniquely to its present value
$\rho_{Q0}=28.8~meV^4$.

Because the observed  $\widetilde{\dub}$ is already close to the
cosmological constant value $-1$, an inverse power potential
requires $\alpha<1$, so that the potential energy {\em always}
dominated the kinetic energy ($x<<y$). These nearly flat
trajectories never track, but "crawl" \cite{Huey} towards their
present values, only because they were initially tuned close to
these values.

\subsection{SUGRA Potentials Now Curve Over Towards Slow Roll}
A good tracker with large basin of attraction must, while tracking,
have small potential curvature.  But, if it is to slow
roll down to $\widetilde{\dub}<-0.78$, after tracking, it must develop a large
potential curvature $\eta$ ($\beta(\phi)$ must decrease, $P(\phi)$
increase), so that $\omq\sim t^P$ grows rapidly at late times
\cite{SWZ}. Such {\em cross-over quintessence} \cite{Cald} is
characterized by $\dub (z)$ reducing in the recent past ($z<0.5$).

The roll-over behavior needed is illustrated in the popular potentials
listed in Table III.  (A longer list of potentials is given in
references \cite{Well,Sahni,Peebles,Starob,Padman}.) For example, the
SUGRA models on the bottom row of Table 1 have minima at
$\varkappa\phi=\sqrt\alpha$, beyond which the curvature $\eta$
increases, allowing $\beta$ to decrease precipitously from $\alpha$ in
the tracking era, to $\mathcal{O}\sim (\alpha-1)^2/(\alpha+1)$ at
present.  After tracking in the background-dominated era, these SUGRA
phase trajectories, summarized on the bottom two rows of Table II and
dotted in Figure 1, curve over towards lower $w_{Q0}$ values, in
marginal agreement with observations, for a large range in $\alpha$
values.

\input{TableIII.tex}

%% file: TableIII.tex
\begingroup
\begin{table*}
\caption{Quintessence potentials which track early, but slow-roll
  now.}
\tiny
\begin{ruledtabular}
   \begin{tabular}{|l|c|r|}
   {\bf Potential $V(\phi)$} &{\bf Theoretical Origin}&{\bf References}\\  \hline
    $M^4[\cos(\phi/f)+1]$&String, M-theory pseudo Nambu-Goldstone light axion&\cite{Frieman2,CDF,CDS,Choi}\\
    $M^{4+\alpha} \phi^{-\alpha} \cdot\exp{\frac{1}{2}(\varkappa\phi)^\beta/2}$&SUGRA, minimum at $(\varkappa\phi)^\beta=2\alpha/\beta$                       &\cite{Bin,Brax,Brax2}      \\
    $M_P^4 [A+(\varkappa\phi-\varkappa\phi_m)^{\alpha}]\exp(-\lambda\varkappa\phi)$&Exponential modified by prefactor, to give local minimum; M-theory &\cite{Albr,Skor}
   \end{tabular}
\end{ruledtabular}
\end{table*}
\endgroup
\normalsize

%% file: Conclusions.tex
\section{CONCLUSION: CANONICAL QUINTESSENCE REQUIRES TWO COSMIC COINCIDENCES}
We have not considered modified gravity, quantum corrections to
classical general relativity, topological defects, non-canonical
scalar fields, a true cosmological constant, nor matter-coupled
quintessence.  But, within canonical quintessence, the
observations allow phase trajectories that are insensitive to initial
conditions, only if the potential's curvature increases rapidly, just
before the present epoch.

Difficult combined supernova, CBR and cosmic shear observations in the
next decades may yet tell whether the smooth energy is static or
dynamic and, if dynamic, whether it is quintessence, k-essence or 
even not driven by a scalar field.  Otherwise, a dynamic smooth energy appears
hardly distinguishable, theoretically and phenomenologically, from the
small cosmological constant it was designed to explain.

The original cosmological coincidence problem was to understand why
the smooth energy density is {\em now} so small, after allowing large
scale structure formation, necessary prerequisites of life and
consciousness. This coincidence problem is now compounded by the
requirements that the quintessence potential energy now be small {\em
and} its curvature now be fast-increasing. In two ways, recent
cosmological observations distinguish the special time in which we
live, in support of weak anthropic reasoning in cosmology.